\documentclass[useAMS,usenatbib,usegraphicx]{mn2e}

\usepackage{graphicx}
\usepackage[latin1]{inputenc}
\usepackage{color}
\usepackage{times}
\usepackage{natbib}
\usepackage{setspace}
\newif\ifAMStwofonts
\AMStwofontstrue
\definecolor{red}{rgb}{1,0.,0.}

\newcommand{\morgana}{{\sc morgana}}
\newcommand{\lcdm}{$\Lambda$CDM }
\newcommand{\lsfr}{\ell_{\rm sfr}}
\newcommand{\msun}{{\rm M}_\odot}

\def\lesssim{\lower.5ex\hbox{$\; \buildrel < \over \sim \;$}}
\def\gtrsim{\lower.5ex\hbox{$\; \buildrel > \over \sim \;$}}
\title[IMF variations in SAMs] {Variations of the Initial Mass
  Function in Semi-Analytical models.}

\author[Fontanot F.]{
  \parbox[t]{\textwidth}{Fabio Fontanot\thanks{E-mail: fontanot@oats.inaf.it}}
    \vspace*{8pt}\\
    INAF - Astronomical Observatory of Trieste, via G.B. Tiepolo 11, I-34143 Trieste, Italy \\
}

\begin{document}
\date{Accepted ... Received ...}

\maketitle

\begin{abstract} 
Deviations from a universal, MW-like, Stellar Initial Mass Function
(IMF) have been reported for distant galaxies, although the physical
reason behind the observed variations is still matter of ongoing
debate. In this paper, we present an exploratory study to assess the
impact of the proposed IMF evolution on the statistical galaxy
properties, as predicted by the Semi-Analytical model of galaxy
formation and evolution \morgana. In particular, we test different
dependencies for the IMF shape, as a function of both model galaxy
properties (such as star formation rate, velocity dispersion or
stellar mass) and environment, and compare the predicted stellar mass
functions and star formation rate functions with reference runs at
fixed IMF. In most cases, \morgana~predictions show deviations of the
order of a few tenths of dex with respect to a run assuming an
Universal Kroupa IMF. Among the proposed IMF variations, an increasing
Top-Heavy IMF at increasing star formation rates has the largest
impact on predicted galaxy properties, while most of the models
assuming an increasing Bottom-Heavy IMF at higher masses/velocity
dispersion lead to variations in galaxy properties that are of the
same order as the uncertainty on the mass and star formation rate
determination. By comparing the predicted galaxy stellar mass
functions, we conclude that the study of the high-mass end can provide
useful constraints to disentangle models assuming an increasing
Top-Heavy IMF in high star forming or Bottom-Heavy IMF in massive
systems.
\end{abstract}
\begin{keywords}
  galaxies:evolution - galaxies: fundamental parameters - galaxies:
  star formation - galaxies: luminosity function, mass function
\end{keywords}

\section{Introduction}\label{sec:intro}

The characterisation of the role played by the different physical
processes acting on baryonic gas in shaping the observed properties of
galaxy populations is still a fundamental goal for modern
astrophysics. Among those mechanisms expected to impact galaxy
evolution, the description of star formation is a key asset for any
theoretical model of galaxy formation and evolution. However, despite
the huge scientific effort in understanding the chain of events
leading to the formation of single and/or clustered stars from an
unstable gas cloud, a number of long-standing problems are still open
\citep[see e.g.][and references herein]{Krumholz14}. In particular,
the definition of the Stellar Initial Mass Function (IMF, hereafter),
which regulates the relative abundance of massive versus low-mass
stars per each stellar generation, has seen considerable recent debate
about possible variation of its shape as a function of galaxy
properties.

Historically, the IMF has been estimated via stellar counts in the
solar neighbourhood and/or the Milky Way (MW), where stellar
populations can be measured with enough accuracy up to low-mass
stars. Different functional representations for the IMF shape in the
MW have been proposed in the literature, from early suggestions of a
single power-law \citep[][S55]{Salpeter55}, to a broken power-law
\citep[][K01]{Kroupa01} or lognormals with power-law tail
\citep[][C03]{Chabrier03}. Relevant uncertainties are connected with
both the brown dwarf regime below 0.08 $\msun$ (where star counts are
expected to decrease, but the sharpness of the decline is poorly
constrained) and to the cut-off at high masses between 100 and 150
$\msun$. It is very difficult to repeat such detailed analysis in
distant galaxies, as only the integrated light of multiple stellar
populations is observationally accessible: therefore the apparent
invariance of the IMF among MW regions\footnote{But see
  \citet{Klessen07} for a possible deviation from an universal IMF in
  the Galactic Centre.} has been usually assumed to be an universal
property of star forming regions.

Theoretical models investigating the origin of the IMF shape from
first principles predict some degree of variability as a function of
the physical conditions of the star forming medium. These include
models trying to describe star formation in a turbulent medium
\citep[see e.g][]{Klessen05, HennebelleChabrier08, Hopkins12}, models
based on the Jeans mass argument \citep[see e.g][]{NarayananDave13}
and models investigating the role of cosmic rays as star formation
regulators \citep{Papadopoulos10}. The claim for a systematic change
of IMF with the mass of parent stellar clusters led to the notion of
integrated galactic IMF \citep[see e.g.][]{KroupaWeidner03}: since the
mass spectrum of stellar clusters in a galaxy is not uniform, this
implies that the integrated galactic IMF is different from that of
individual clusters.

From an observational point of view, several authors reported a
flattening of the IMF at increasing star formation rates (SFR) in late
type galaxies \citep{HoverstenGlazebrook08, Gunawardhana11}, on the
base of multi-colour photometry. The study of unresolved stellar
populations in external galaxies also shows indications for a
non-universal IMF: \citet{Cappellari12} compared mass-to-light ratios
derived from stellar kinematics to the predictions of stellar
population synthesis models, finding systematic variations of about a
factor of two with respect to a MW-like IMF. Similar results have been
obtained spectroscopically using spectral features sensible to the
stellar effective temperature and surface gravity
\citep{vanDokkumConroy11, ConroyvanDokkum12}, with the data suggesting
increasingly bottom-heavy IMFs for galaxies with larger velocity
dispersions. It is not clear to what extent the \citet{Cappellari12}
and \citet{ConroyvanDokkum12} results agree among themselves
\citep[see e.g.][]{Smith14} and/or are in tension with the
increasingly Top-Heavy IMFs in high star forming galaxies \citep[see,
  e.g.][]{NarayananDave13}. So far no clear evidence for a redshift
evolution of the IMF shape has been found: \citet{ShettyCappellari14}
studied a sample of 68 $0.7<z<0.9$ field galaxies, and they found an
average normalisation of the IMF in massive galaxies consistent with
S55 slope, with a substantial scatter.

Any IMF change will impact both the predictions of theoretical models
and the reconstruction of galaxy properties from multi-wavelength
photometry, i.e. through Spectral Energy Distribution (SED) fitting
techniques. In the latter approach, it is still possible to shift
between models at fixed IMF by means of correction factors, but the
same approach does not hold for the predictions of theoretical models
\citep[see e.g.][]{DeLuciaBlaizot07}, as a different IMF has a larger
effect than simply changing the amount of baryonic mass locked in
long-lived stars\footnote{It is worth stressing that applying a rigid
  shift to the predictions of a theoretical model calibrated with a
  given IMF to compare it with data extrapolated using a different IMF
  is still a valid approach (e.g when several models are compared to
  the same dataset), even if the correct approach would require to
  include the shift in the data.}. The pictures became more
complicated if we allow the IMF to vary along galaxy evolution: in
this case it is not possible to recover galaxy physical properties
from algorithms comparing observed to synthetic photometry, without
any prior knowledge of the relation between star formation history and
IMF shape. Also in the case of an algorithm relaxing the universal IMF
hypothesis and best-fitting also the shape of the IMF, like in the
\citet{vanDokkumConroy12} approach, the resulting shape represents the
IMF of the dominant stellar population, and/or an integrated mean
value along the whole galaxy lifetime.

In this paper, we present an exploratory study of the effect of IMF
variations on the predictions of Semi-Analytical models of galaxy
formation and evolution \citep[SAMs,see e.g.][for a review of this
  approach]{Baugh06}. To this aim, we will explicitly include in the
model different parametrization of IMF variation, both theoretically
and observationally grounded. We will focus on the SAM parameters that
directly depend on the IMF definition, while keeping all other model
parameters fixed, with the aim of highlighting the impact of a varying
IMF shape on the predicted physical properties of model
galaxies. Earlier attempts to include different IMF shapes in the SAM
framework have been already presented in the literature
\citep{Baugh05}, but the novelty of our approach lies in the wide
range of possible IMF variations and dependencies considered. In the
following, there will be no attempt to compare model predictions to
physical quantities derived under the hypothesis of universal IMF
shape, as we consider such a comparison misleading. Moreover, we do
not attempt a comparison with direct photometry either, as we want to
highlight the modifications on the basic physical properties of
galaxies and avoid any possible additional uncertainty due to the
choice of a given simple stellar population library. This paper is
organised as follows. In Sec.~\ref{sec:models} we describe the IMF
variable models we implement on the the SAM model \morgana; in
Sec.~\ref{sec:results}, we present our results, while in
Sec.~\ref{sec:final} we give our conclusions.

\section{Semi-Analytical model}\label{sec:models}

Modern theoretical models of galaxy formation and evolution assume
that Dark Matter (DM) haloes are the privileged sites for galaxy
formation, driven by a complex network of physical processes including
at least (but not only) the cooling of baryonic gas, the onset of star
formation and the various feedbacks effects associated with the death
of massive stars and the accretion of cold gas on to Super-Massive
Black Holes. In an attempt to overcome our limited knowledge of these
key physical mechanisms, SAMs describe them using simple mathematical
prescriptions (either physically or observationally motivated) and
then follow the time evolution of the different galaxy components
(bulge, disc and halo) and gas phases (stars, cold gas, hot
gas). Quantitative comparisons between the predictions of different
SAMs have been extensively discussed in the literature, showing the
relative strength of this approach, which provides, at least, a
coherent picture of galaxy evolution in the \lcdm concordance
cosmological model \citep[see e.g.][]{Fontanot09b, Fontanot12a}.  This
level of coherence was not expected {\it a priori}, given the
different implementations for the relevant physical processes assumed
by the different groups developing SAMs, and represents another facet
of the relevant degeneracies involved in the definition of the
parameter space associated with SAMs themselves.

In this paper, we consider predictions from the MOdel for the Rise of
Galaxies aNd Agns (\morgana): we refer the reader to \citet{Monaco07}
for a full description of the model features, including the modelling
of cooling, star formation, black hole accretion and feedback
processes, and to \citet{LoFaro09} for the definition of the most
recent calibration of the model, based on the C03 IMF. In \morgana,
the shape of the IMF determines the number of supernovae (SNe) per
unit stellar mass formed ($f_{\rm sn}$) which regulates the strength
of SNe feedback and the matter/energy exchange between the cold and
hot gas phases (see, e.g., \citealt{Fontanot13} for a comparison of
different feedback schemes in SAMs). For each stellar generation, a
fraction of the baryonic material involved in the star formation
process is eventually given back to the hot gas phase through stellar
winds, mass loss and SNe ejecta. In \morgana, this {\it returned
  fraction} ($R$) regulates the mass flow between the stellar phase
and the hot gas reservoir of the halo, setting the amount of recycled
gas. Moreover, this material is supposed to be enriched in metals,
formed in the stellar interiors as a consequence of stellar
evolution. The pollution of the primordial gas with these metals has a
strong impact on the predicted cooling rates, given the dependency of
the cooling function on the hot gas metallicity
\citep{SutherlandDopita93}. As the IMF controls the relative abundance
of massive versus low-mass stars in star forming events, its shape is
thus fundamental for computing both the amount of baryons locked in
long-lived stars and the amount of metal spread in the Inter-Stellar
and Inter-Galactic Media.

In \morgana, the process of star formation is treated separately in
the disc and bulge, due to the different physical conditions in the
two galaxy components \citep[see][for a detailed comparison of the
  \morgana~implementation of star formation with respect to other
  SAMs]{Fontanot13}. In particular, the cold gas associated with the
bulge (due to mergers, disc instabilities and/or direct infall through
the cooling flow) is converted into stars on a timescale which is
usually shorter with respect to the ``quiescent'' star formation in
discs \citep{Monaco07}. Although we consider separately IMF variations
taking place in the disc and bulge components, it is worth stressing
that most of the stellar mass ending up in the bulge component is
formed in discs (then transferred via mergers and/or discs
instabilities), with only a limited contribution from {\it in-situ}
star formation.

We run all \morgana~models considered in this study on the same merger
tree ensemble, extracted from a 200 {\rm Mpc} box, with concordance
cosmology ($\Omega_0=0.24$, $\Omega_\Lambda=0.76$, $h=0.72$,
$\sigma_8=0.8$, $n_{\rm sp}=0.96$). This cosmological realisation has
been obtained using the Lagrangian code {\sc pinocchio}
\citep{Monaco02} with N=1000$^3$ particles: the particle mass in this
run is thus $2.84 \times 10^8 \msun$ , with the smallest used dark
matter halo being $1.42 \times 10^10 \msun$ (50 particles) and the
smallest resolved progenitor $2.84 \times 10^9 \msun$ (10
particles). 

In detail, we define the IMF as the number of star counts at a given
mass and we choose its normalisation $A$ by requiring that the mass
integral equals 1 $\msun$:

\begin{equation}\label{eq:imf_i}
A \int_{m_{\rm low}}^{m_{\rm up}} m \, \phi(m) \, dm = 1
\end{equation}

\noindent
where $m_{\rm low}=0.08 \msun$ and $m_{\rm up}=120 \msun$ represent
our assumed integration limits; $f_{\rm sn}$ is thus:

\begin{equation}
f_{\rm sn} = A \int_{M_{\rm SN}}^{m_{\rm up}} \phi(m) \, dm
\end{equation}
\noindent
where $M_{\rm SN} = 8 \, \msun$ represents the minimum initial mass
for a star bound to end up its evolution as a Type II Supernova. In
the useful hypothesis of the Instantaneous Recycling Approximation
(IRA), we assume that stellar lifetime are negligible with respect to
the integration timestep of the model and we can thus estimate $R$ as:

\begin{equation}
R = A \int_{m_{\rm low}}^{m_{\rm up}} [m-r(m)] \, \phi(m) \, dm
\end{equation}
\noindent
where $r(m)$ represents the mass fraction locked into stellar remnants
(i.e. long-lived stars, white dwarfs, neutron stars and black holes)
at the end of stellar evolution as a function of initial stellar
mass. Finally, we compute the stellar yield $y_z$, i.e. the mass
fraction given to the surrounding gas in the form of newly synthesised
metals:

\begin{equation}\label{eq:imf_f}
y_z = \frac{A}{1-R} \int_{m_{\rm low}}^{m_{\rm up}} m \, y(m) \, \phi(m) \, dm 
\end{equation}
\noindent
where $y(m)$ represents the stellar yield of a star of initial mass
$m$, during its whole lifetime. In this paper, we consider both $y_z$
and $y(m)$ as {\it effective} yields, referring to all synthesised
metals $Z$, i.e. we choose not to follow the chemical evolution of
individual elements. Our choice is motivated by the fact that this
paper mainly focus on the global statistical properties of galaxy
populations, and the study of detailed chemical patterns are beyond
the goal of this work; however we recognise that chemical evolution is
possibly a key discriminant between different models of IMF variation
\citep[see e.g.][]{Gargiulo14} and we devote its study to future
developments.

In this paper, we adopt the $r(m)$ and $y(m)$ values estimated by
\citet[][in particular their Tables~4,~5,~6]{Maeder92}. In that work,
grids of evolutionary stellar models ranging from 1 to 120 $\msun$
were used to derive chemical yields and remnant masses as a function
of the initial stellar mass. We applied this results to stellar masses
larger than 1 $\msun$ (a linear interpolation is used between the grid
points): we thus assumed that lower mass stars do not contribute
significantly to both $R$ and $y_z$. Results for models with both a
solar $Z=0.02$ and a lower $Z=0.0001$ metallicity have been provided:
we choose between the two options on the basis of the predicted
metallicity of the cold star forming gas in model galaxies. The use of
just two tabulated values for metallicity has little impact on our
results. In fact, all \morgana~runs considered in this work predict
cold gas metallicities around or above the the solar value for most
model galaxies at $z \lesssim 2-3$. In practice, we use the
low-metallicity tables only at higher redshifts and at the low-end of
the mass function. We explicitly test that our main conclusions do not
change, using the \citet{Maeder92} tables relative to solar abundances
for all galaxies.

\subsection{IMF variations}
\begin{figure}
  \centerline{ \includegraphics[width=9cm]{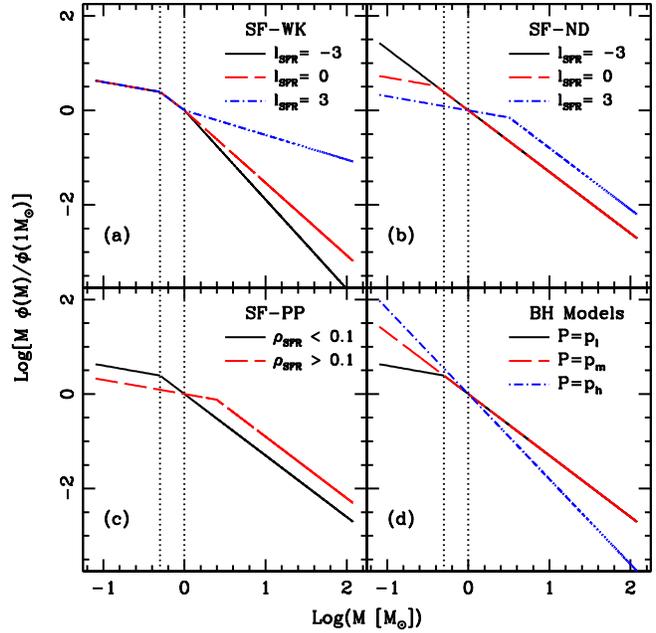} }
  \caption{Predicted IMF shape variations for the models considered in
    this work (see text for the definition and units of labelled
    quantities). In each panel, IMFs are normalised to the 1 $\msun$
    value. Vertical dotted lines mark the position of the two break
    masses ($m^\prime_1$, $m^\prime_2$) in the canonical, MW-derived,
    IMF.}\label{fig:imfcomp}
\end{figure}

As a generalised shape for the IMF we adopted the multi-power-law, 5
parameters, K01 IMF:

\begin{equation}
\phi(m) =  \left\{
\begin{array}{ll}
(\frac{m}{m_{\rm low}})^{\alpha_1} & m_{\rm low} \le m < m_1 \\
(\frac{m_1}{m_{\rm low}})^{\alpha_1} (\frac{m}{m_1})^{\alpha_2} & m_1 \le m < m_2 \\
(\frac{m_1}{m_{\rm low}})^{\alpha_1} (\frac{m_2}{m_1})^{\alpha_2} (\frac{m}{m_2})^{\alpha_3} & m_2 \le m \le m_{\rm up} \\
\end{array}
\right.
\label{eq:kroimf}
\end{equation}

\noindent
This shape provides the required versatility to study IMF changes in
well defined mass intervals, as a function of galaxy properties; the
canonical, MW-derived, IMF is defined by the primed parameter choice
$(\alpha^\prime_1, \alpha^\prime_2, \alpha^\prime_3) =
(-1.30,-2.30,-2.30)$ for the exponents and $(m^\prime_1, m^\prime_2) =
(0.5 \msun, 1 \msun)$ for the break stellar masses.,

From a technical point of view, we integrate Eq.~\ref{eq:imf_i}
to~\ref{eq:imf_f} to compute $f_{\rm SN}$, $R$ and $y_z$ on an
appropriate table covering the desired range of IMF variations, and
then, for each stellar generation, select the entry closer to the
required input. We consider several models for IMF variation that we
summarise in the following, while we provide a visual comparison of
the sensible range of IMF shapes in Fig:~\ref{fig:imfcomp}. In this
figure, we mark the position of the canonical break masses
$m^\prime_1$ and $m^\prime_2$ and we normalise each IMF to its value
at 1 $\msun$, to highlight the relative contribution of high- and
low-mass stars.

\begin{itemize}
{\item {\bf Models SF}. A number of recent paper has pointed out a
  possible dependence of the IMF shape on the SFR of individual
  galaxies \citep[see e.g.][]{WeidnerKroupa06, Gunawardhana11}. In
  this class we include different parametrization as suggested by
  different groups.

\begin{itemize}
{\item {\bf SF-WK model} is based on the integrated galactic IMF model
  of \citet{WeidnerKroupa06}: we fixed all parameters as in the
  canonical IMF, but $\alpha_3$, which we assume to be a piecewise
  function of $\lsfr = \log({\rm SFR})$:

\begin{displaymath}        
\alpha_3 =  \left\{        
\begin{array}{lll}          
-3.187 & \lsfr < -4 \\       
0.300 \, (\lsfr+3) -2.887 & -4 \le \lsfr < -3 \\  
0.175 \, (\lsfr+2) -2.712 & -3 \le \lsfr < -2 \\  
0.090 \, (\lsfr-0.25) -2.51 & -2 \le \lsfr < 0.25 \\   
0.36\, \lsfr -2.6 &   \lsfr \ge 0.25 \\        
\end{array}                
\right.                    
\end{displaymath}          
\noindent
where the functional forms have been estimated from comparison with
\citet[][their Fig.~2]{Weidner13}. The SFR dependence implies that
high-SF regions have IMF that are more Top-Heavy than moderate-to-low
SF regions (Fig.~\ref{fig:imfcomp}, Panel (a)).}

{\item {\bf SF-ND model} follows the Jean mass argument for giant
  molecular clouds evolution as detailed in
  \citet{NarayananDave13}. They use high-resolution hydrodynamical
  simulations to calibrate a parametrization of the IMF break (or
  characteristic) mass as a function of the galaxy SFR:

\begin{displaymath}
m_1(SFR) = 0.5 \, \left( \frac{SFR}{2 M_\odot \, {\rm yr}^{-1}}
\right)^{0.3}
\end{displaymath}
\noindent
As for the other parameter in Eq.~\ref{eq:kroimf}, we fixed $\alpha_2
= \alpha^\prime_1$ if $m_1(SFR)>m^\prime_2$, $\alpha_2 =
\alpha^\prime_3$ otherwise. In practice, we reduce the K01 form to a
double power shape (Fig.~\ref{fig:imfcomp}, Panel (b)). This
parametrization corresponds to a Top-Heavy IMF for high star formation
events, while providing, at the same time, a bottom-heavy\footnote{It
  is worth stressing, that given the parametrization assumed in
  \citet{NarayananDave13} and in our work, the maximum level of
  bottom-heaviness allowed correspond to a S55 IMF.} IMF in low-star
forming galaxies, the discriminant being set at the MW value.}

{\item {\bf SF-PP model} moves from the theoretical work of
  \citet{Papadopoulos10}. In this approach, the shape of the IMF is
  explained by assuming that the thermal and ionization state of dense
  clouds is determined by Cosmic Rays (which are able to penetrate
  deeply into molecular clouds) rather than optical-to-UV photons. For
  a reasonable range of cosmic ray energy densities this translates
  into a narrow range of possible temperatures and ionization states
  for dense gas clouds, and thus to an almost invariant IMF. However,
  at high cosmic rate energy densities, typical of extreme
  environments like compact starburst, relevant deviations from an
  universal IMF are expected, due to the rather different thermal
  state of the dense gas and to the different fragmentation of the
  cloud. This model thus predicts two IMF regimes
  (Fig.~\ref{fig:imfcomp}, Panel (c)): a MW-like IMF is assumed
  overall, but in compact starbursts, which are associated with a
  Top-Heavy IMF. The latter environments are identified by assuming a
  critical value of the SFR-density $\Sigma_{\rm SFR}>0.1 \msun$
  yr$^{-1}$ Kpc$^{-2}$, roughly about 100 times the MW value
  (Papadopoulos, private communication). Above this threshold we then
  assume a sudden change of the IMF break mass $m_1$ from a typical MW
  $\sim 0.5 \msun$ value to $\sim 2.5 \msun$ \citep{Papadopoulos11},
  and we fix $\alpha_2 = \alpha^\prime_1$ as in model SF-ND. }
\end{itemize}
}
\begin{table}
  \caption{IMF variation: BH models}
  \label{tab:models}
  \centering
  \begin{tabular}{ccccc}
    \hline
    Model & $\mathcal{P}$ & $p_l$ & $p_m$ & $p_h$ \\
    \hline
    BH-SG1 & $\log(\sigma_B)$ & 1.9 & 2.2 & 2.5 \\
    BH-SG2 & $\log(\sigma_{\rm cold})$ & 1.9 & 2.2 & 2.5 \\
    BH-SG3 & $\log(V_H)$ & 1.9 & 2.2 & 2.5 \\
    \hline
    BH-MSa & $\log({\rm M_B}/\msun; {\rm M_D}/\msun)$  & 10.75  & 11.25  & 12  \\
    BH-MSb & $\log({\rm M_B}/\msun; {\rm M_D}/\msun)$  & 9  & 10.5  & 12  \\
    \hline
    BH-MTa & $\log({\rm M_\star}/\msun)$  & 10.75  & 12.25  & 12  \\
    BH-MTb & $\log({\rm M_\star}/\msun)$  & 9  & 10.5  & 12  \\
    \hline
    BH-DMa & $\log({\rm M_H}/\msun)$  & 12  & 12.5  & 14  \\
    BH-DMb & $\log({\rm M_H}/\msun)$  & 10  & 12  & 14  \\
    \hline
    BH-z0a & $\log({\rm M_{\rm H0}}/\msun)$ & 12  & 12.5  & 14  \\
    BH-z0b & $\log({\rm M_{\rm H0}}/\msun)$ & 10  & 12  & 14  \\
    \hline
  \end{tabular}
\end{table}

{\item {\bf Models BH} This category includes different realisations,
  inspired from the suggestion of an increasingly Bottom-Heavy IMF at
  increasing velocity dispersion or galaxy stellar mass. We mimic
  these behaviours by considering a shape evolution of the IMF as
  described below. We define $\mathcal{P}$ as the galaxy property we
  assume the IMF is dependent on, and $p_l<p_m<p_h$ as three typical
  values of $\mathcal{P}$ describing the relative shape evolution.

\begin{displaymath}
\alpha_1 =  \left\{
\begin{array}{ll}
-1.30 & \mathcal{P} \le p_l \\
-2.35 + \frac{1.05 (\mathcal{P}-p_m)}{p_l-p_m} & p_l < \mathcal{P} \le p_m\\
\alpha_3 & \mathcal{P} > p_m \\
\end{array} 
\right.
\end{displaymath}

\begin{displaymath}
\alpha_2 =  \left\{
\begin{array}{ll}
-2.35 & \mathcal{P} \le p_m \\
\alpha_3 & \mathcal{P} > p_m \\
\end{array} 
\right.
\end{displaymath}

\begin{displaymath}
\alpha_3 =  \left\{
\begin{array}{ll}
-2.35 & \mathcal{P} \le p_m \\
-2.80 + \frac{0.45 (\mathcal{P}-p_h)}{p_m-p_h} & \mathcal{P} > p_m \\
-2.80 & \mathcal{P} \ge p_h \\
\end{array} 
\right.
\end{displaymath}

\noindent
Qualitatively (Fig.~\ref{fig:imfcomp}, Panel (d)), between $p_l$ and
$p_m$, only $\alpha_1$ is evolving, transforming the IMF from the K01
to the S55 shape; for $\mathcal{P}>p_m$ the IMF is considered as a
single power-law with an increasingly Bottom-Heavy IMF with a maximal
slope of $-2.80$ \citep[see e.g.][]{Cappellari12} at $p_h$. We test
different recipes varying the key galaxy property (see
Tab.~\ref{tab:models} for a summary). We also test different choices
for the $p$ values, given the somehow arbitrary definition of these
parameters. One $p$ set (``a'' models) is chosen by requiring that
$p_l$ is set around the MW-like scale, while $p_h$ represents the
scale of massive galaxies and $p_m$ is an intermediate scale. We then
define an alternative $p$ setting by moving $p_l$ and $p_m$ to lower
values, allowing IMF variations starting from lower-mass systems
(``b'' models).

\begin{itemize}
{\item {\bf BH-SG models} assume that the IMF shape depends on the
  velocity dispersion of model galaxies, i.e. they closely resemble
  the results presented by \citet[][their Fig.~4]{Conroy13}.
  Different definition for this quantity in SAMs have been considered:
  the velocity dispersion of the bulge $\sigma_B=$ (BH-SG1), the
  velocity dispersion $\sigma_{\rm cold}$ of the cold clouds in the
  bulge (BH-SG2, defined as in Sec.~7.1 of \citealt{Monaco07}) and the
  circular velocity $V_H$ of the host halo (as a proxy for the
  velocity dispersion of the galaxy, BH-SG3). In BH-SG1 and BH-SG2,
  the changes in IMF shape are allowed only in the bulge component
  (i.e. the IMF is K01 invariant in the disc component), while in
  BH-SG3 the variation is assumed in both the bulge and the disc.}

{\item {\bf BH-MS models} assume that the IMF variation depends on the
  mass of the star forming galaxy, either by component (i.e. the mass
  of the disc $M_D$ and the bulge $M_B$ separately), or total
  ($M_\star=M_B+M_D$).}

{\item {\bf BH-DM models} assume that the IMF variation is mostly linked
  with the environment, through the halo mass $M_H$.}

{\item {\bf BH-z0 models} link the IMF variation to $M_{\rm H0}$,
  defined as the $z=0$ parent halo mass for central galaxies and the
  mass of the hosting substructure right before DM merging for
  satellites. The rationale behind this class is the difficulty of a
  straightforward comparison between \citet{ConroyvanDokkum12} or
  \citet{Cappellari12} results and the predictions of our models, due
  to our assumption of IMF variations along the galaxy lifetime, while
  their approach still assumes that each galaxy has a fixed, but not
  universal, IMF. By using $M_{\rm H0}$, as a proxy of the final mass
  of the galaxy under consideration, we can artificially force a fixed
  IMF for each model galaxy. It is worth stressing that this model is
  clearly idealised, as there is no basis to the hypothesis that star
  formation events at a given redshift should depend on the $z=0$
  environment.}
\end{itemize}
} 

\end{itemize}

These models represent a basic set of IMF variations: they include
dependencies on integrated galaxy properties, such as SFR, $\sigma_B$,
stellar masses of bulge and disc (Models SF, BH-SG1-2, BH-MS). They
also consider IMF variations closely related to the Large Scale
Structure, such as possible environmental dependencies on the
properties of parent DM halo (Models BH-SG3, BH-DM and BH-z0). Despite
the lack of a clear theoretical link between a local process like star
formation and the properties of the Large Scale Structure as defined
by the cosmological model, those former models are designed to explore
the possible influence of the cosmological evolution in setting up the
properties of the baryonic content of the haloes. Moreover, the models
we consider include both Top-Heavy and Bottom-Heavy IMF variations,
thus allowing us to study both trends.

\section{Results}\label{sec:results}
\begin{figure*}
  \centerline{ \includegraphics[width=18cm]{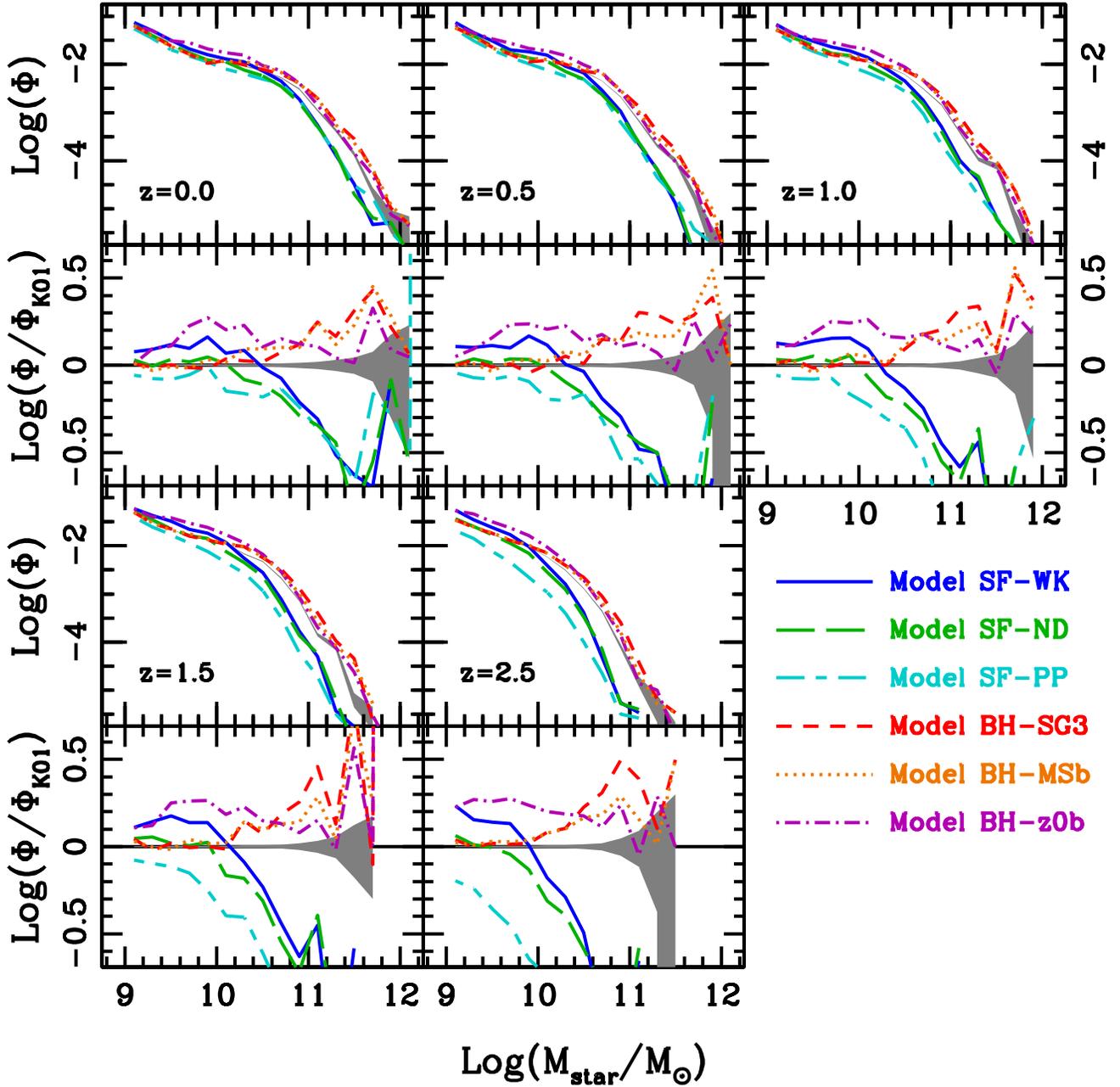} }
  \caption{Redshift evolution of the predicted stellar mass function
    for different IMF models. In each panel, the grey shading
    represents the statistical error associated with the determination
    of the stellar mass function in our cosmological box, computed on
    the reference K01 IMF run. Line types and colours correspond to
    the prediction of different models, as labelled.}\label{fig:mfevo}
\end{figure*}
\begin{figure*}
  \centerline{ \includegraphics[width=18cm]{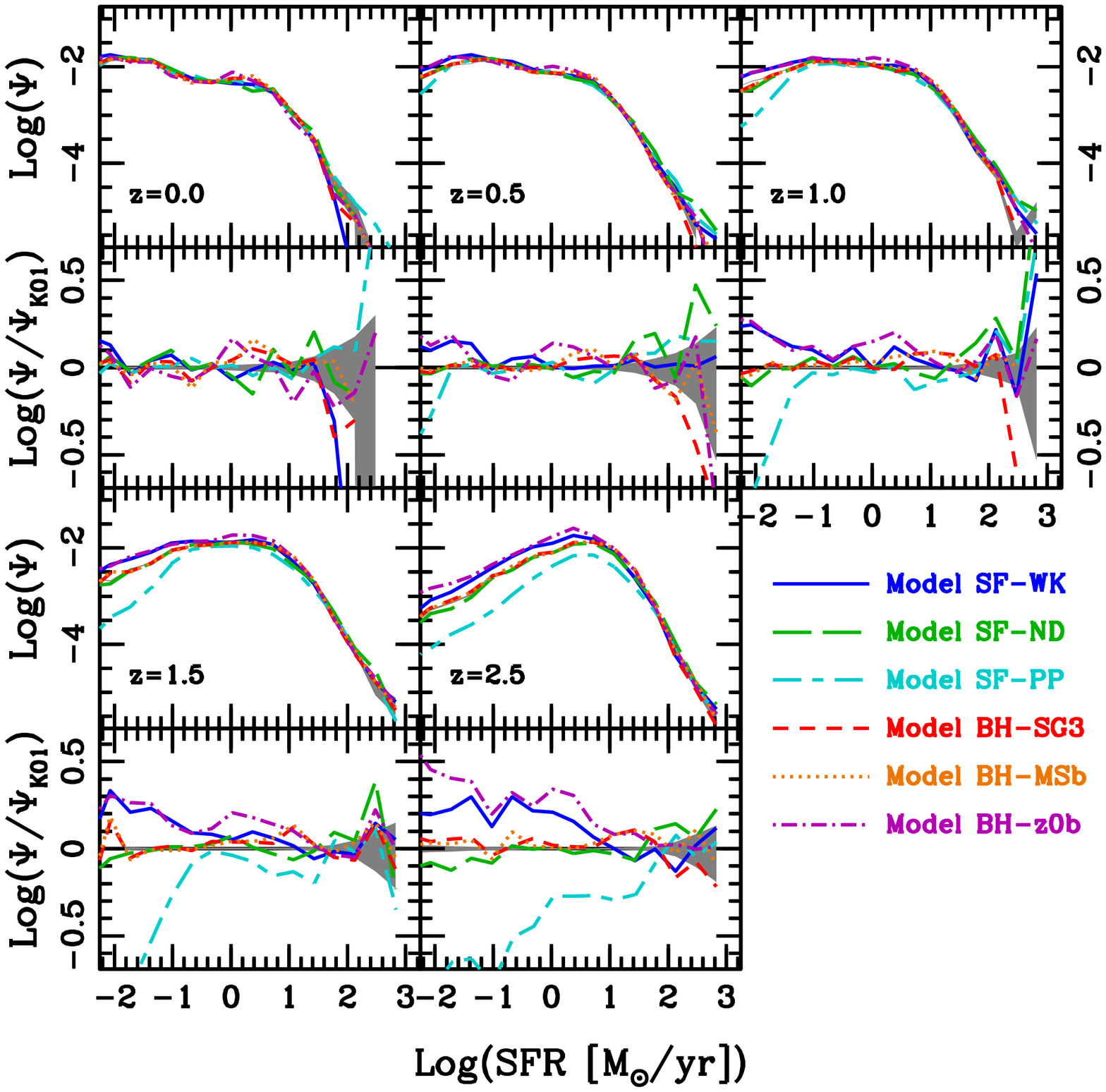} }
  \caption{Redshift evolution of the predicted SFR function for
    different IMF models. In each panel, shaded areas and model
    predictions are shown with the same lines types and colours as in
    Fig.~\ref{fig:mfevo}.}\label{fig:sfrfevo}
\end{figure*}

We then compare the redshift evolution of the galaxy stellar mass
function (Fig.~\ref{fig:mfevo}) and SFR function
(Fig.~\ref{fig:sfrfevo}) as predicted by our models. We define a
reference model with a fixed (universal) K01 IMF and we use the
corresponding predictions to estimate the statistical error associated
with the stellar mass/SFR functions in the simulated cosmological
volume (poissonian errorbars shown as a shaded area in
Fig.~\ref{fig:mfevo} and Fig.~\ref{fig:sfrfevo}). We recall that our
reference \morgana~realisation is calibrated on C03 IMF. We test that
the differences in model predictions moving from a C03 to a K01 IMF,
at fixed SAM parameter space, are negligible, with respect to the
variations implied by the IMF models presented in this paper. At each
redshift in Fig.~\ref{fig:mfevo} and~\ref{fig:sfrfevo}, we also show
the ratio between the stellar mass/SFR function in a given model and
the reference K01 realisation. For a sake of clarity, in both figures
we shown only 3 BH models, chosen as representative of the whole
class.

SF models provide large deviations ($\sim$ 0.5 dex or larger) from the
reference K01 IMF run in the stellar mass function, in particular, at
$M_\star>10^{10} M_\odot$. Those differences grow larger as redshift
increases and high star formation events became more important, while
the corresponding stellar populations dominate the mass budget of
model galaxies. The effect is particularly dramatic in the SF-PP
model, which shows the largest deviation from the reference MF at all
mass scales. SF-PP is also the model showing the largest differences
in the SFR function, in particular at the low-SFR end and at $z>1$,
while the SF-WK and SF-ND predictions are consistent with the
reference model at most SFR levels. For all SF models, some
interesting deviations (of the order of a few tenth of dex as in
stellar mass function case) are seen in the most extreme starburst and
at higher redshifts (which are the environments where we expect the
different gas recycling factors implied by the different IMF
variations to have the larger impact), but in general the SF model
have little effect on the SFR function shape.

\begin{figure}
  \centerline{ \includegraphics[width=9cm]{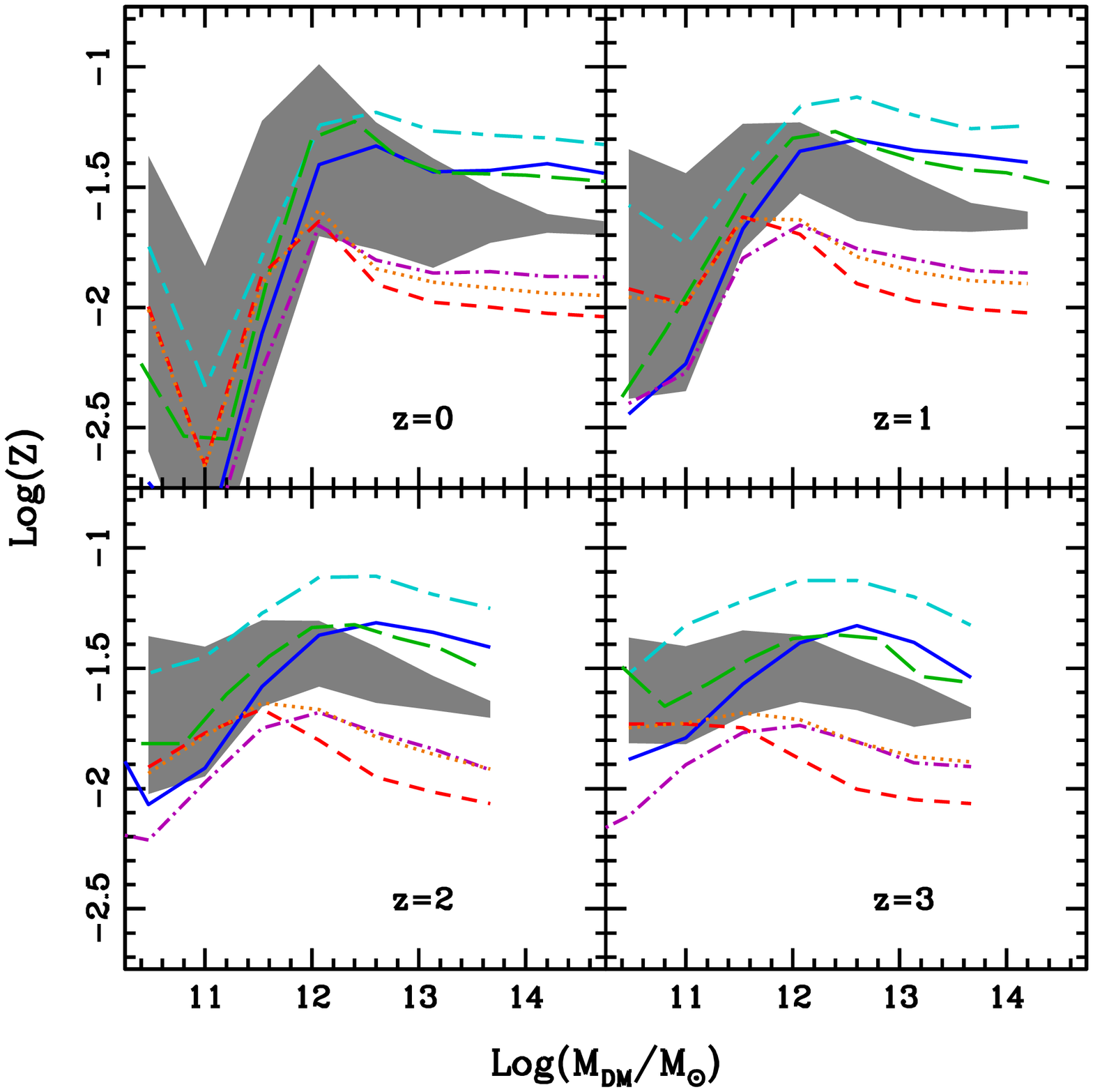} }
  \caption{Redshift evolution of the hot gas metallicities as a
    function of DM halo masses. In each panel, model predictions are
    shown with the same lines types and colours as in
    Fig.~\ref{fig:mfevo}, while the shaded area represents the
    1-$\sigma$ scatter in the mean relation for the reference, K01
    IMF, model.}\label{fig:zhot}
\end{figure}

In order to deepen this crucial point we consider in
Fig.~\ref{fig:zhot} the metallicity of the hot gas as a function of
the parent DM halo mass, in order to estimate the effect of the IMF
variation on cooling rates. The shaded area refers to the predictions
of the reference K01 IMF run (1-$\sigma$ scatter around the mean
value), while other models are marked with the same colour coding as in
Fig.~\ref{fig:mfevo}. SF runs predict enhanced metallicities of the
hot gas, due to the larger $y_z$ from their (mostly) Top-Heavy IMF:
this in turn translates into larger cooling rates, which balance the
increased feedback strength (larger $f_{\rm sn}$), leading to the small
differences in the distribution of the instantaneous SFR.

As far as the BH models are considered, our results show milder
effects with respect to the SF runs, with small deviations from the
reference K01 IMF run in most cases. In particular, ``a'' models do
not differ from the reference MFs by more than $\sim 0.1-0.2$ dex over
the whole mass range (thus we choose not to show them in the figures),
while in ``b'' models larger differences, of the order of
$\sim0.3-0.5$ dex, are seen at the high mass end. In terms of SFR
function, BH-MS and BH-DM models predictions are completely consistent
with the reference run. The overall behaviour of BH-MS and BH-DM model
is due to the fact that in all these models relevant IMF variations
take place only at high mass scales (both stellar or DM): galaxies
reaching those stages of their evolution are usually experiencing
decreasing (if not quenched, see e.g. \citealt{Kimm08}) star formation
histories, thus only a small fraction of stars are formed with a
Bottom-Heavy IMF. This is particularly severe in ``a'' models, where
the range of IMF variations is reduced by construction. The model
showing the largest deviations from the reference run is BH-SG3, while
both BH-SG1 and BH-SG2 predict negligible deviations at most masses,
given the fact that only a small fraction of the stars are forming in
the bulge component, where IMF variations are allowed.

These results imply that variations of the IMF should be effective in
the standard ``quiescent'' star formation mode (associated with
discs), in order to support a mass-dependent IMF scenario, as expected
given the limited contribution of star formation in bulges to the
total stellar mass of model galaxies. Moreover, IMF variations should
also be effective in relatively low-mass galaxies: this further point
is tested in BH-z0 models, which is explicitly designed to link the
level of IMF variation to the final parent halo/substructure mass,
taken as a proxy to the final galaxy mass. This run, despite
unrealistic, is built up to maximise the level of bottom-heaviness
associated with a BH approach, and enhance any evolutionary effect
connected to a mass-dependent Bottom-Heavy IMF, since, at all epochs,
each model galaxy uses an IMF set by its $z=0$ properties (according
to the BH-MS model). It is thus interesting to see that BH-z0 models
show differences of the order of $\sim0.2-0.3$ dex at most mass scales
and redshift, still compatible with the intrinsic uncertainty in mass
reconstruction from observed photometry. BH-z0 models show larger
deviations with respect to the predictions of other BH models in terms
of SFR function, especially at $z>1$. However, also for these models
deviations are of a few tenths of dex at most.

\section{Discussion \& Conclusions}\label{sec:final}
In this paper, we present an exploratory study on the impact of IMF
variations in Semi-analytical models of galaxy formation and
evolution. By assuming IMF variations to depend either on the physical
properties of star forming galaxies (SFR, velocity dispersion, stellar
mass) or on the properties of the Large Scale Structure (parent halo
mass), we run different SAM realisations. Several definitions for the
relevant quantities have been considered, and different
parametrizations, broadly inspired by the observational results. Our
results thus test the SAM robustness against the hypothesis of
moderate IMF variations

In general, typical deviations in the galaxy mass function
predictions, with respect to the reference K01 IMF run, are of the
order of few tenths of dex over a wide redshift range, thus of the
same order as the uncertainty due to different choices of universal
(fixed) IMF (i.e. between K01 and S55), while smaller effects are seen
in the star formation rate functions. In BH models, these differences
are comparable to the scatter between the predictions of different
SAMs (once they are calibrated against the same reference sample, see
e.g. \citealt{Fontanot09b, Fontanot12c}), and due to the different
parametrizations of the relevant physical processes. Only models
assuming an increasing Top-Heavy IMF in high SFR environments predict
deviations from the reference model larger than both the statistical
and modelling uncertainties. The model showing the largest deviations,
both in terms of stellar masses and SFRs, from the reference K01 IMF
run is SF-PP. In order to get a better understanding of the processes
leading to these differences we run a couple of additional runs,
varying the key hypothesis of this model, namely the value of the
$\Sigma_{\rm SFR}$ threshold and/or allowing the IMF variation only in
a single galaxy component (i.e. the disc or the bulge). Our tests
highlight the fact that its peculiar behaviour arises as a consequence
of our treatment of star formation in discs. In both SF-WK and SF-ND,
low mass galaxies are affected only weakly by IMF variations, as a
consequence of their mostly MW-like SFRs; \morgana, however, predicts
SFR surface densities in discs higher than the SF-PP threshold, at
most galactic scales. In fact, if the IMF change is allowed only in
the bulge component, SF-PP predicts smaller deviations, and a low-mass
end completely consistent to the reference K01 run, given the limited
amount of stars directly forming in the bulge component. While these
results are indeed interesting and possibly suggesting a deep revision
of our understanding of galaxy evolution, few caveats are worth
discussing. The main limitation of our modelling of the
\citet{Papadopoulos10} approach lies in the treatment of galaxy
components, i.e., galactic discs are considered as single entities,
with no attempt to model density and SFR gradients. Thus the SFR
surface densities entering the definition of SF-PP model are mean
densities, integrated over the whole disc. Alternative descriptions,
as those based on the statistics of individual star forming cloud,
and/or on local properties of star forming regions will be of great
interest\citep[see e.g.][for an attempt to overcome this
  limitation]{Fu10}.
\begin{figure}
  \centerline{ \includegraphics[width=9cm]{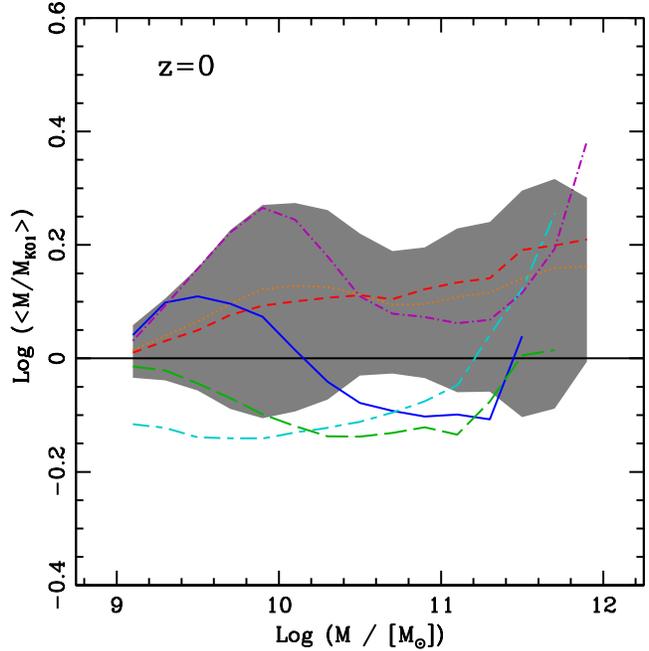} }
  \caption{Mass deviations from reference K01 IMF model. Model
    predictions are shown with the same lines types and colours as in
    Fig.~\ref{fig:mfevo}, as labelled. Shaded area represents the
    1-$\sigma$ scatter in BH-SG3 model (see text for more
    details).}\label{fig:deltamag}
\end{figure}

Moreover, the relatively small effect seen in most of our models
implies that the complex interplay between the physical properties
modelled in SAMs has the net effect of smoothing out some of the
differences due to the description of instantaneous star formation in
a variable IMF framework. In fact, by changing the IMF shape, we are
not only changing the fraction of baryons locked in long lived stars
and remnants per each stellar generation, but also the amount of
recycled baryons and the energetics of the multiphase gas. We
explicitly show the level of self-regulation at place in our SAM,
while discussing the evolution of the SFR function and we interpret
its stability, in most of our models, as the result of the interplay
between the stronger (weaker) stellar feedback and the enhanced
(depressed) cooling rates, the latter due to the higher (lower) metal
enrichment of the hot gas phase. We thus conclude that the main effect
of the IMF change in \morgana~is set by the amount of baryons locked
in low mass stars (responsible for the stellar mass of model
galaxies). As a general caveat, we remind the reader that the
distribution of metals into the different gas phases (both hot and
cold) is extremely sensible to the (poorly constrained) details of the
modelling of gas reheating and ejection \citep[see][for a review of
  the different implementations of these processes in
  SAMs]{Fontanot13}. Additional tests with other SAMs are thus
required to assess if the self-regulation of the SFR is a general
property in the SAM framework.

In order to better understand the effect of IMF variations on the
evolutionary tracks of individual galaxies, we compare the $z=0$
stellar masses predicted by each model to those predicted by the
reference K01 IMF run, on an object-by-object basis. In
Fig.~\ref{fig:deltamag}, we show the mean mass ratios $\Delta M =
\log(<M/M_{\rm K01}>)$ for the same models considered in
Fig.~\ref{fig:mfevo}, as a function of model galaxy mass. We also
report with a shaded area the 1-$\sigma$ scatter in the distribution
for the BH-MSb model, which is representative. Overall, mean $\Delta M
= \pm 0.2$dex: Top-(Bottom-)Heavy model have in general negative
(positive) mean deviations, as expected, but the substantial scatter
implies that individual model galaxies might have deviations in
opposite directions. Moreover, mean trends are not monotonic: this
again shows that the impact of IMF variations is different at
different mass scales. A qualitative comparison with \citet{Conroy13}
is possible: while the mean $\Delta M$ are clearly lower, model
galaxies with individual $\Delta M$ as high as their results are
typically less than 2-$\sigma$ outliers for most of the BH models. It
is worth stressing that the sources entering the \citet{Conroy13}
analysis have a peculiar selection function and it is not clear if
they are representative of the whole galaxy population.

By comparing models assuming an increasing Bottom- or Top-Heavy IMF at
increasing stellar mass or SFR, we notice that the two classes provide
predictions, which are clearly different and mutually exclusive at the
high-mass end of the mass function, with the resulting space densities
for massive galaxies being systematically higher or lower than the
reference run predictions (although the separation is not that clear
by considering the mass differences in single objects, as shown in
Fig.~\ref{fig:deltamag}). We can thus conclude that this mass range is
the most promising for breaking the degeneracies between the different
proposed IMF variations (such as in \citealt{Cappellari12}), but it is
important to keep in mind that these results also depend on our choice
for the IMF variation rules, which prefer by construction larger
variations in more massive systems. At smaller stellar masses the
situation looks more confused, as a larger fraction of stars in these
objects is born from a reference K01 IMF, and its difficult to
disentangle the predictions of different models, with only the SF-PP
model predicting a decrease of the space density of low-mass galaxies
at increasing redshifts. It is also worth noting that in most cases,
the predicted deviations of our IMF variable models from the reference
run depend on both stellar mass and redshift, i.e. variable IMFs have
a differential effect on the shape of the MFs. This may have relevant
consequences in interpreting galaxy evolution from observed
photometry, i.e. by means of SED fitted physical quantities and in the
context of the so-called {\it downsizing} trends \citep{Fontanot09b}.

As a final remark, we want to recall the reader that despite the
reference \morgana~realisation is calibrated on a C03 IMF, in the
spirit of this work, which is not aimed at comparing data and models,
but at exploring the effect of IMF variations on SAM predictions, we
do not try any recalibration of the reference model: with our choice
we thus focus on the effect of IMF variations only. A complementary
approach would rely on the recalibration of each realisation on
similar reference dataset: however this approach is not currently
feasible. In fact, on one side, we cannot use any dataset including
galaxy physical properties reconstructed by means of SED fitting
algorithms: as we already discussed in the introduction, systematic
IMF variations as a function of stellar (parent halo) mass, SFR or
redshift represent a critical problem for any algorithm trying to
recover galaxy physical parameters from observed photometry, as most
of these algorithms would be sensible (at most) only to the dominant
stellar component. This is coherent with the substantial intrinsic
scatter found by \citet{ShettyCappellari14} when comparing dynamically
derived mass-to-light ratios to the predictions of stellar population
models at fixed (S55) IMF. A possible way out of this problem for this
class of algorithms would then require the self-consistent fit of the
most reliable star formation history for each galaxy (and the
corresponding IMF evolution as well), but this choice increases the
number of fitted parameters and the level of degeneracies, possibly
reducing the significance of the estimated physical
quantities. Therefore, under the hypothesis of a variable IMF, it is
not possible to include the statistical distribution of galaxy
physical properties (such as the stellar mass function) in the
calibration sample for theoretical models. A more solid calibration of
such models would then require the direct comparison with the
photometric properties of galaxy populations like the multi-wavelength
luminosity functions or number counts. This step implies a
straightforward evolution of spectrophotometric codes, allowing them
to mix simple stellar populations characterised by (almost) arbitrary
shaped IMFs in the SAM framework. We left this project to future
developments.

\section*{Acknowledgements}
I warmly thank Pierluigi Monaco and Gabriella De Lucia for discussion
and comments, and Padelis Papadopoulos for useful clarifications about
his model. I also want to thank the anonymous referee for a
constructive report that helped improving the quality of the paper. I
acknowledge financial contribution from the grants PRIN MIUR 2009
``The Intergalactic Medium as a probe of the growth of cosmic
structures'' and PRIN INAF 2010 ``From the dawn of galaxy formation''

\bibliographystyle{mn2e}
\bibliography{fontanot}

\end{document}